\documentclass[a4paper,11pt]{article}
\usepackage{jinstpub} 
\usepackage{lineno}
\usepackage{subcaption}
\usepackage{pgfplots}
\pgfplotsset{compat=1.18} 
\usepgfplotslibrary{groupplots}
\usepackage{circledsteps}

\PassOptionsToPackage{RGB}{xcolor}
\usepackage{xcolor}
\definecolor{KITgreen}     {RGB}{0,150,130}
\definecolor{KITwhite}        {RGB}{255,255,255}
\definecolor{KITdarkblue} {RGB}{0,45,76}
\definecolor{KITiceblue}    {RGB}{168,185,196}
\definecolor{KITicegray}  {RGB}{218,225,230}
\definecolor{KITpinegreen}{RGB}{0,90,80}
\definecolor{KITcyan}      {RGB}{35,161,224}
\definecolor{KITblue}     {RGB}{12,83,126}
\definecolor{KITlightgreen}{RGB}{140,182,60}
\definecolor{KITforestgreen}     {RGB}{39,103,56}
\definecolor{KITblack}     {RGB}{0,0,0}
\definecolor{KITpurple}    {RGB}{163,16,124}
\definecolor{KITorange}    {RGB}{223,155,27}
\definecolor{KITyellow}    {RGB}{252,229,0}
\definecolor{KITred}       {RGB}{211,0,21}
\definecolor{KITbrown}     {RGB}{167,130,46}

\proceeding{23rd International Workshop on Advanced Computing and Analysis Techniques in Physics Research\\
September 8th-12th, 2025\\
Hamburg, Germany}

\title{\boldmath Real-Time Stream Compaction for Sparse Machine Learning on FPGAs}

\author[a,1]{M. Neu,\note{Corresponding author.}}
\author[b]{I. Haide,}
\author[b]{T. Ferber,}
\author[a]{J. Becker}
\affiliation[a]{Institute for Information Processing Technology, Karlsruhe Institute of Technology,\\Engesserstraße 5, 76131 Karlsruhe, Germany}
\affiliation[b]{Institute of Experimental Particle Physics, Karlsruhe Institute of Technology,\\Wolfgang-Gaede-Straße 1, 76131 Karlsruhe, Germany}

\emailAdd{marc.neu@kit.edu}

\abstract{Machine learning algorithms are being used more frequently in the {first-level triggers in collider experiments}, with Graph Neural Networks pushing the hardware requirements of FPGA-based triggers beyond the current state of the art.
To meet the stringent demands of high-throughput and low-latency environments, we propose a concept for latency-optimized preprocessing of sparse sensor data, enabling efficient GNN hardware acceleration by removing dynamic input sparsity.
Our approach rearranges data coming from a large number of First-In-First-Out interfaces, typically sensor frontends, to a smaller number of FIFO interfaces connected to a machine learning hardware accelerator.
In order to achieve high throughput while minimizing the hardware utilization, we developed a hierarchical sparsity compression pipeline optimized for FPGAs.
We implemented our concept in the Chisel design language as an open-source hardware generator. For demonstration, we implemented one configuration of our module as preprocessing stage in a GNN-based first-level trigger for the {Electromagnetic Calorimeter inside the Belle~II detector}. Additionally we evaluate latency, throughput, resource utilization, and scalability for a wide range of parameters, to enable broader use for {other large scale scientific experiments}.}

\keywords{Sparsity Compression, Machine Learning, Graph Neural Networks, Field Programmable Gate Array, Stream Compaction, Data Movement, Trigger Systems}

\newacronym{daq}{DAQ}{Data Acquisition}
\newacronym{l1}{TRG}{Level-1 Trigger}
\newacronym{hlt}{HLT}{High-Level Trigger}
\newacronym{fee}{FEE}{Front-End Electronics}
\newacronym{grl}{GRL}{Global Reconstruction Logic}
\newacronym{gdl}{GDL}{Global Decision Logic}
\newacronym{fpga}{FPGA}{Field-Programmable Gate Array}
\newacronym{cdc}{CDC}{Central Drift Chamber}
\newacronym{klm}{KLM}{K-Long and Muon detector}
\newacronym{ecl}{ECL}{Electromagnetic Calorimeter}
\newacronym{adc}{ADC}{Analog-to-Digital Converter}
\newacronym{tdc}{TDC}{Time-to-Digital Converter}

\newacronym{gnn}{GNN}{Graph Neural Network}
\newacronym{mva}{MVA}{Multi-Variate Analysis}
\newacronym{mlp}{MLP}{Multi-Layer Perceptron}
\newacronym{mc}{MC}{Monte Carlo}

\newacronym{lut}{LUT}{Look-Up Table}
\newacronym{ff}{FF}{Flip-Flop}
\newacronym{dsp}{DSP}{Digital Signal Processing}
\newacronym{pe}{PE}{Processing Element}
\newacronym{fifo}{FIFO}{First-In-First-Out buffer}
\glsdisablehyper
\notoc

\begin{document}
\maketitle
\flushbottom
\glsresetall
\section{Introduction}
\label{sec:intro}
\begin{figure}[b]
    \centering
    \includegraphics[width=0.5\linewidth]{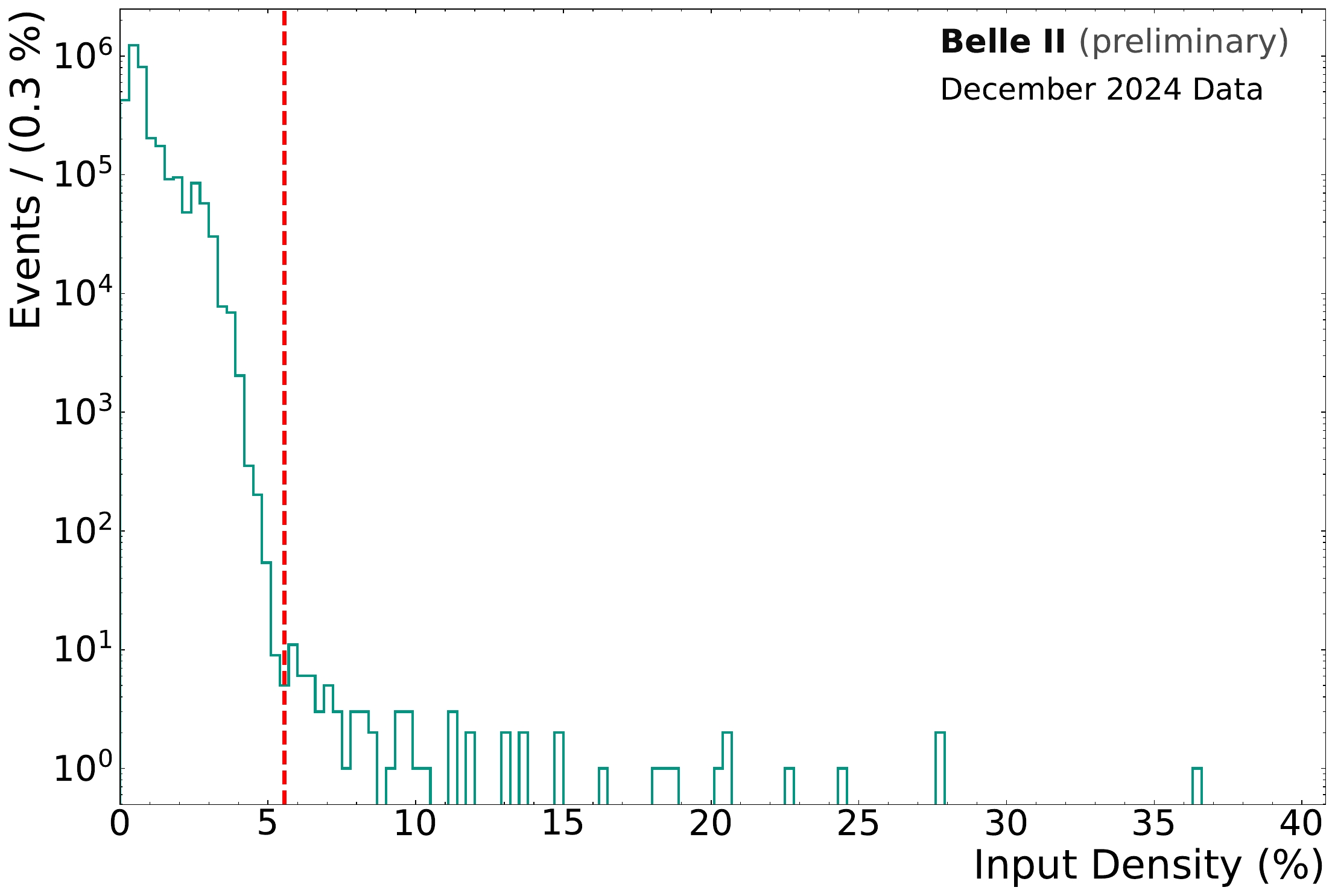}
    \caption{Histogram of input data density per event as received by the GNN-ETM module in the first-level trigger system at Belle~II. The data density is defined as the fraction of non-zero data values out of all possible data values in an event.}
    \label{fig:sparsity}
\end{figure}

Machine learning algorithms are increasingly adopted in first-level triggers in collider experiments~\cite{zipper:2024,Bahr:2024dzg,Liu:2026iup}.
In this context, Graph Neural Networks (GNNs) require efficient deployment strategies on FPGA-based trigger hardware to satisfy the stringent latency and throughput constraint~\cite{deiana:2022,shlomi:2021,abadal:2022}.
While there is a range of conceptual studies on the deployment of GNNs for first-level trigger systems~\cite{Dittmeier:2025nlh,Que:2025mhf,elabd:2022}, end-to-end pipelines required for the actual deployment in the collider experiment have not yet been considered.\\
Recently, dynamic GNNs such as GraVNet \cite{wemmer:2023,qasim:2019} have attracted growing interest for first-level trigger applications. However, their $O(N^2)$ computational complexity with respect to the number of inputs presents a significant challenge for low-latency, high-throughput environments. Furthermore, particle physics data are typically very sparse. As an illustration, \autoref{fig:sparsity} shows the input density observed in the Belle~II electromagnetic calorimeter (ECL) during data taking at the highest ever recorded instantaneous luminosity.\\
This work addresses the question of how input sparsity can be exploited to enable the efficient computation of dynamic GNNs in first-level trigger systems. In particular, our approach must satisfy several key challenges: (1) hard real-time requirements, as trigger decisions must be produced deterministically; (2) minimal latency overhead, given that the total trigger latency budget is in the order of microseconds; and (3) high throughput demands in the order of 10 million detector snapshots per second.\\
To overcome these challenges, we make the following contributions: (1) we present an open-source hardware generator implemented in Chisel; (2) we  evaluate the performance of the proposed sparsity compression module across a range of configurations, quantifying its impact on latency, throughput, and resource utilization; (3) we demonstrate the feasibility of our approach through deployment within the GNN-ETM, a newly developed hardware trigger module in the Belle~II first-level trigger system.

\section{Background}
\label{sec:background}

\begin{figure}
    \centering
    \includegraphics[width=\linewidth]{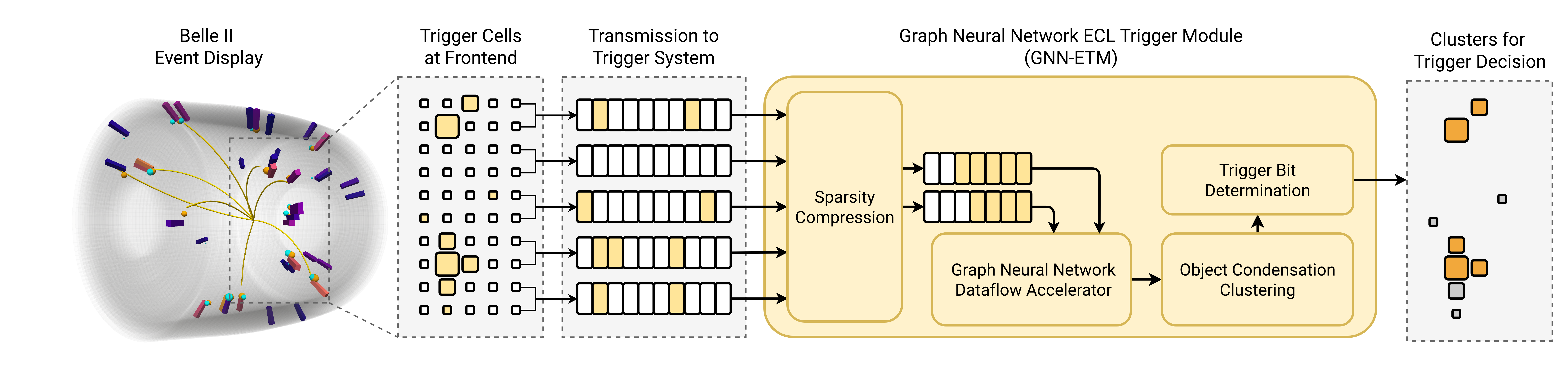}
    \caption{Overview of the Graph Neural Network ECL Trigger Module inside the Belle II first-level trigger system. This diagram is simplified to give an conceptual overview of the sparsity compression module inside the system.}
    \label{fig:introduction}
\end{figure}

The Belle~II Experiment \cite{Belle-II:2010dht} at the SuperKEKB electron-positron collider in Tsukuba, Japan, studies flavour and dark sector physics at the $\Upsilon$(4S) resonance energy of \SI{10.58}{\giga\electronvolt}. 
At a current world-record instantaneous luminosity of \SI{5.1e34}{\centi\metre^{-2}\second^{-1}} at SuperKEKB, the beam-induced backgrounds lead to higher detector occupation and thus higher throughput demands on the Belle~II data acquisition (DAQ) system~\cite{Yamada:2015xjy}.
The DAQ system supports a maximum event readout rate of $30\,\text{kHz}$, which is limited by the available bandwidth at the detector frontends. To reduce the computational load on the DAQ system, a first-level trigger system is employed~\cite{Lai:2025gac}. This trigger operates synchronously with the detector frontend readout at $127.216\,\text{MHz}$. Detector snapshots are processed strictly sequentially. To prevent buffer overflows in the DAQ readout system, the first-level trigger must satisfy a hard realtime latency deadline of 4.4\,µs at all times. \\
The Belle~II first-level trigger system is composed of dedicated subtriggers for participating subdetector. The electromagnetic calorimeter (ECL) first-level trigger identifies energy depositions~\cite{kim:2017}.
Signals from the 8736 thallium-doped cesium iodide crystals are first summed in the analog domain using $4\times4$ crystal groups. A waveform fit is then performed to extract the signal amplitude and timing relative to the highest energetic TC. The resulting preprocessed signals, referred to as Trigger Cells (TCs), are uniquely identified by their positions within the detector.\\
In the current ECL first-level trigger system, TCs are clustered using two approaches: (1) an isolated cluster number logic implemented in the Isolated Cluster Network ECL Trigger Module (ICN-ETM)~\cite{cheon:2002}; and (2) a  GNN–based method implemented in the Graph Neural Network ECL Trigger Module (GNN-ETM)~\cite{haide:2026}. \autoref{fig:introduction} shows an overview of the GNN-ETM architecture. For each $125\,\text{ns}$ timing window, 576 TCs are transmitted to the GNN-ETM via parallel gigabit transceivers (GTs). To enable efficient GNN inference, a sparsity compression module is required, reducing the computational load by extracting non-zero data TCs.

\section{Real-Time Stream Compaction}
\label{sec:concept}

\begin{figure}
    \centering
    \includegraphics[width=0.5\linewidth]{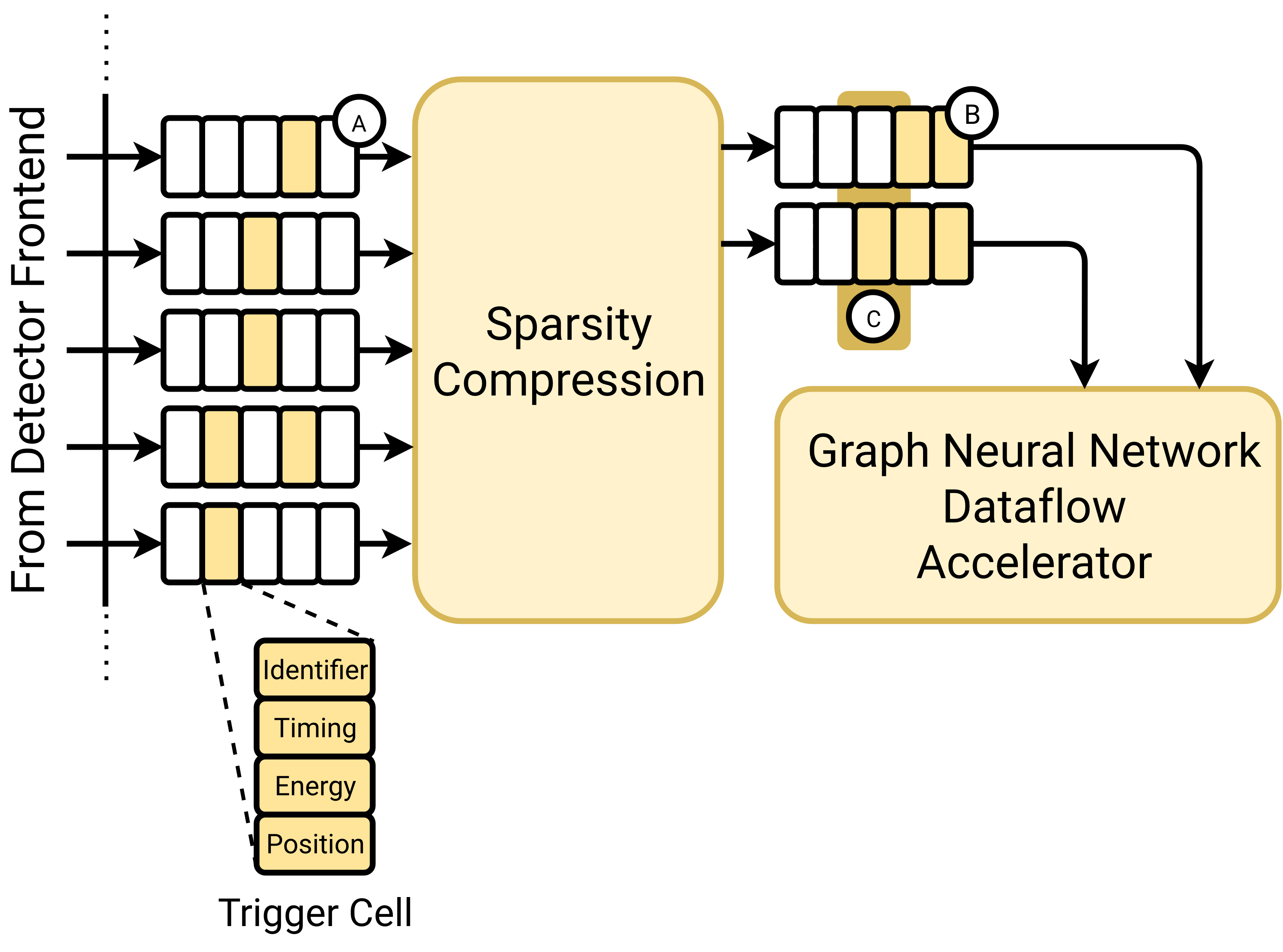}
    \caption{Typical application of the sparsity compression module, illustrated using the Belle II ECL as an example. Sparse data are provided by the detector frontend and compressed for subsequent processing in a dataflow accelerator.}
    \label{fig:concept:overview}
\end{figure}

\begin{figure}
    \centering
    \begin{subfigure}{.50\textwidth}
        \includegraphics[width=\linewidth]{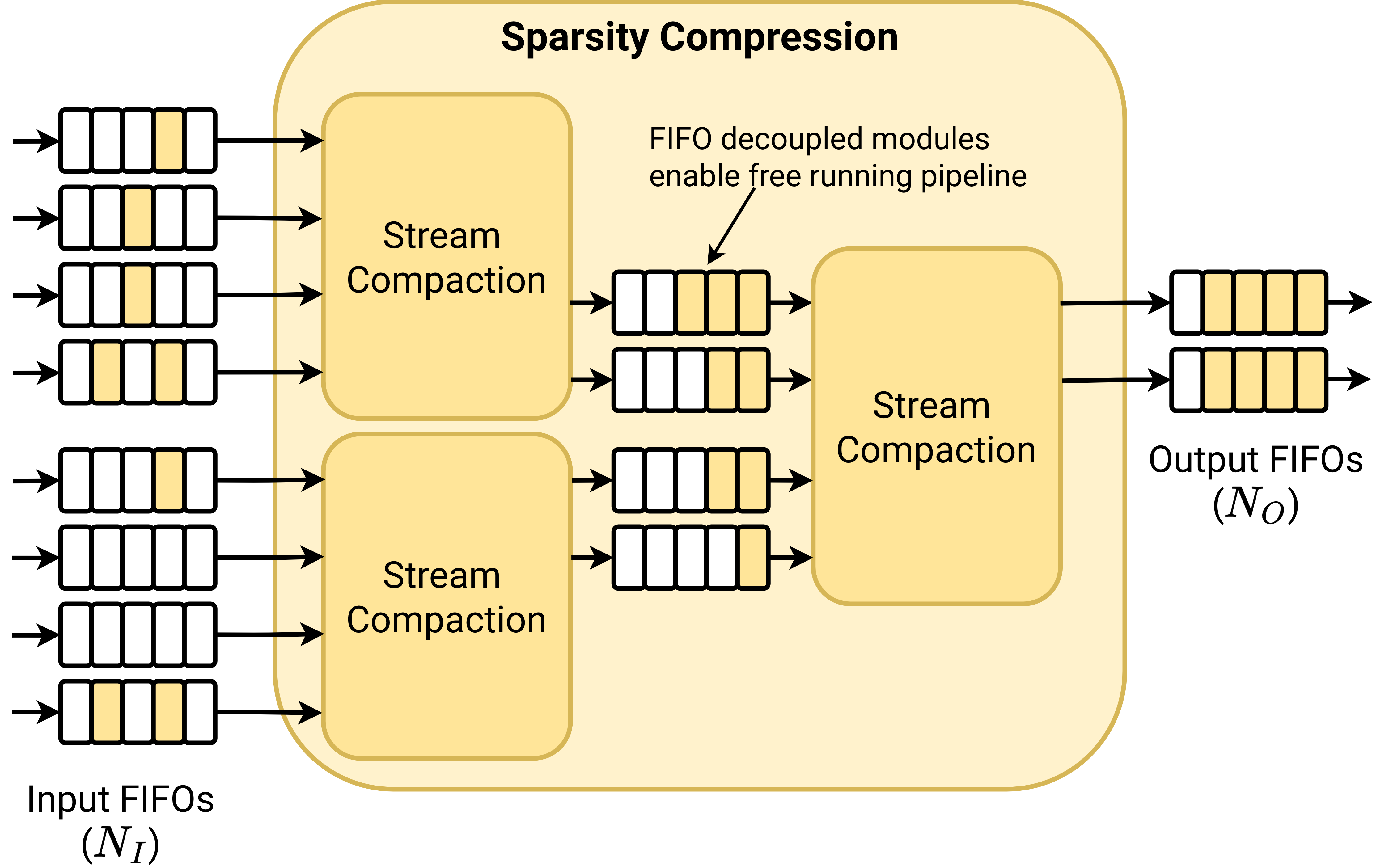}
        \caption{Sparsity Compression}
        \label{fig:concept:sparsitycompression}
    \end{subfigure}
    \hspace{1cm}
    \begin{subfigure}{.28\textwidth}
        \includegraphics[width=\linewidth]{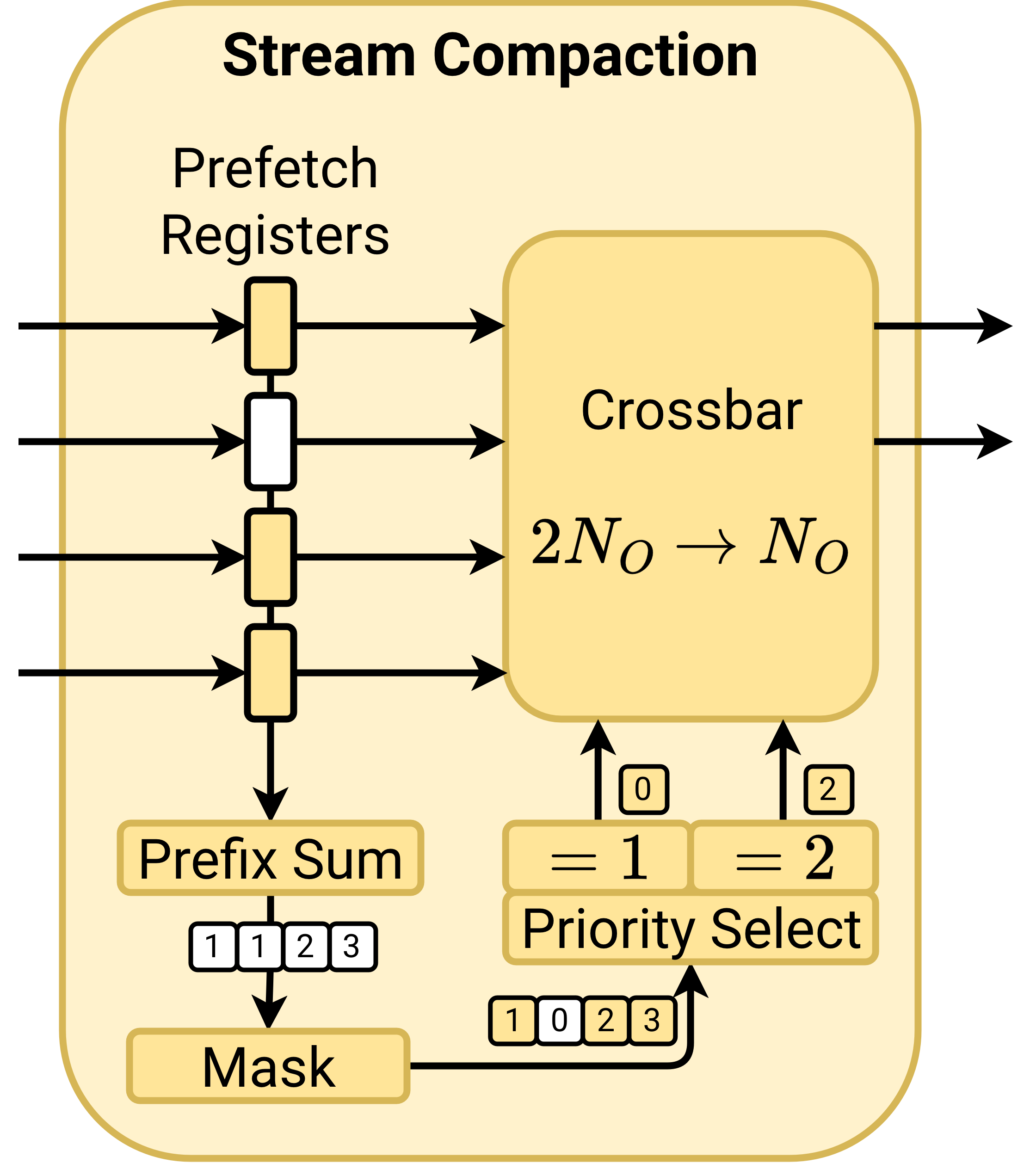}
        \caption{Stream Compaction}
        \label{fig:concept:streamcompaction}
    \end{subfigure}
    \caption{Conceptual block-level diagram of our sparsity compression hardware module. In the exemplary case, the following values are chosen: $N_I = 5$, $N_O = 2$, and $D =5$.}
    \label{fig:concept}

\end{figure}

In literature, sparsity-aware stream compression has been considered for database applications~\cite{papaphilippou:2025,oge:2013} as well as for near sensor bandwidth compression of time-series data for scientific applications~\cite{yoshii:2023}.
However, these work focus on minimizing bandwidth between platforms instead of providing a prepossessing stage for neural network inference. 
In our approach, we reduce a large number of parallel interfaces with sparse data to a smaller number of parallel interface, with the intention of densifying the streams on the same compute platform, maximizing subsequent utilization in pipeline stages.\\
An conceptual top-level system diagram  of our approach is  shown in \autoref{fig:concept:overview}.
The sparsity compression module receives $N_I$ first-in-first-out (FIFO) interfaces as input ports and provides $N_O$ FIFO interfaces as output ports. In every clock cycle, each input port receives one data element. Data elements from up to $D$ consecutive clock cycles are processed as a common window. Non-zero data elements are yellow, whereas empty data elements are white. For clarity, one such data element could be a TC as described in \autoref{sec:background}.\\
The sparsity compression module fulfills two functions, intra- and inter-stream alignment. For the former, the sparse input data \Circled{A} is rearranged into dense streams \Circled{B}. For the later, the dense streams are balanced \Circled{C}. In the following, we will explain how we achieve this functionality.\\
The sparsity compression module depicted in \autoref{fig:concept:sparsitycompression} is composed of a stream compression modules. Each cell inside the hierarchical tree compresses $2\cdot N_O$ input ports into $N_O$ output ports. All cells are connected via FIFO interfaces in a tree topology. The number of stages depends on the ratio between the input and output ports of the stream compaction module.\\
The stream compaction module depicted in \autoref{fig:concept:streamcompaction} is central module of our sparsity compression approach. The three pipeline stages work as as follows: (1) data elements are loaded into prefetch registers; (2) a bitmask is generated from the valid signal, a prefix sum over all valid bits is calculated and applied to the input mask. Two daisy-chained priority encoder select the addresses of the first two non-zero data element; and (3) based on the previously calculated addresses, the crossbar is configured to forward the data elements to the correct output ports.\\
Our approach achieves a number of useful properties: (1) the stream compaction module achieves a deterministic latency as described in \autoref{eq:x}; (2) by specifying the depth of the window $D$ statically, the stall-free pipeline enables 100\% utilization; (3) routeability of the design on FPGA is retained, as crossbars size only scales with the ratio of the module ports.
\begin{equation}
L = 3 \left\lceil \log_2 \left ( 1 + \frac{N_I}{N_O} \right ) \right\rceil f_{\text{sys}}, \quad N_I > N_O
\label{eq:x}
\end{equation}

\section{Implementation and Evaluation}
\label{sec:evaluation}

We implemented our sparsity compression concept as a hardware generator in the Chisel design language~\cite{bachrach:2012}. Configurable parameters are the the number of input ports $N_I$, the number of output ports $N_O$, the bitwidth per port as well as the maximum FIFO buffer size required (depth). The generated streaming ports are AXI-Stream~\cite{ARM:IHI0051B} compliant and support backpressure via the ready-valid handshake .

To evaluate our hardware generator, we implement nine different configurations with varying $N_I$, and $N_O$ parameters. For the remaining parameters, we choose a fixed bitwidth of $32\,\text{bit}$ per port and a depth of 16 entries. We generate the Verilog modules using Chisel~3.6.0 . The functionality of generated modules is verified in register-transfer-level simulation using ModelSim~2023.4~\cite{modelsim:2023:4}. We also verify that the total latency of the module matches our calculation in \autoref{eq:x}.
For all modules, we perform synthesis, place and route in AMD~Vivado~2024 using default directives~\cite{vivado:2024:2}. As FPGA target, we choose the AMD~Ultrascale~XCVU190 as it is the largest available FPGA on the Universal Trigger Board 4 used in the first-level trigger at the Belle~II Experiment.

\autoref{fig:utilization} depicts the utilization of system ressources, namely lookup tables~(LUTs), Registers~(FFs), and Configurable Logic Blocks~(CLBs) for the different configurations. Block RAMs and Digital Signal Processors are not utilized for our design.
It can be seen that the number of LUTs and FFs scale linear with the number of inputs ports. In contrast, the number of output ports scales worse than linearly. Overall the number of occupied CLBs closely matches the number of occupied LUTs, which suggest a good routeability of the design on FPGAs. 

\begin{figure}[ht]
    \centering

    \begin{tikzpicture}
        \small
        \begin{axis}[
            xlabel={Number of Input Ports $N_I$},
            ylabel={System Frequency in MHz},
            width=.7\linewidth,
            height=6cm,
            grid=major,
            ymin=150,
            ymax=550,
            mark size=1pt,
            mark options={scale=2,solid},
            line width=1pt,
            minor y tick num=10,
            ytick={200,300,400,500},
            yticklabels={200,300,400,500},
            xtick={64,128,256},
            xticklabels={64,128,256},
            legend style={
                at={(1.05,0.0)},
                anchor=south west}
        ]
        
        \addplot[ color=KITblue,dotted,mark=*,very thick] coordinates {
            (64,500) (128,483) (256,500) };
        \addlegendentry{$N_O = 2 $}
    
        \addplot[ color=KITpinegreen,dotted,mark=square*,very thick] coordinates {
            (64,470.8) (128,447) (256,387) };
        \addlegendentry{$N_O = 4$}
    
        \addplot[ color=KITpurple,dotted,mark=triangle*,very thick] coordinates {
            (64,301) (128,281) (256,277) 
            };
        \addlegendentry{$N_O = 8$}
    
        \node[draw=none,align=center,inner sep=1pt] (label1) at  (rel axis cs: 0.25,0.55) {{Deployed in GNN-ETM}};
        \node[draw,align=center,shape=circle, color=KITred, inner sep=2pt] (label2) at  (64,500) {~~~};
        \draw[->] (label1) -- (label2);
    
        \end{axis}
    \end{tikzpicture}
    \caption{System frequency for various module configurations after out-of-context synthesis, place and route with AMD~Vivado~2024.2 for the AMD~Ultrascale~XCVU190. One instance of the design is deployed in the GNN-ETM at Belle~II on the same FPGA.}
    \label{fig:frequency}
\end{figure}
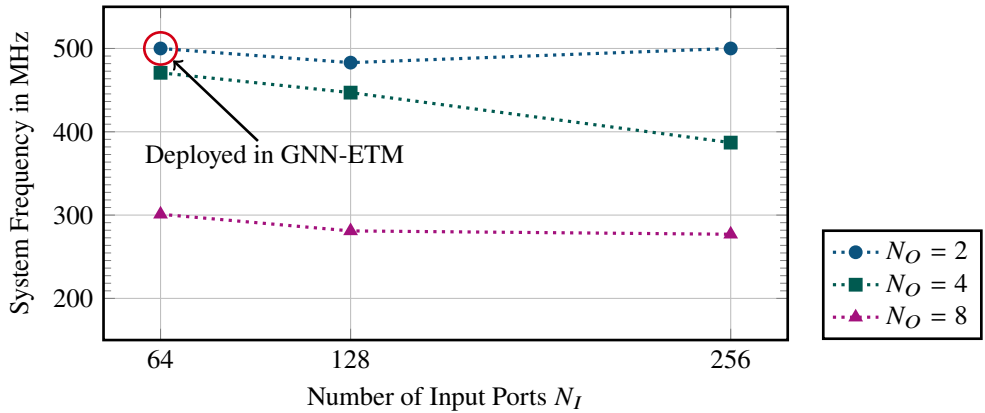

\autoref{fig:frequency} depicts the reported system frequency after out-of-context implementation. To estimate this frequency, we constrain the design to $500\,\text{MHz}$ and derive the achievable frequency from the observed negative timing slack.
For design entities with two output ports, the requested frequency is achieved for two designs. Increasing the number of output ports lowers the realized system frequency down to $277\,\text{MHz}$.
We explain this result in the increased size of the crossbar in the stream compaction module. 
\section{Discussion and Conclusion}
\label{sec:conclusion}

We presented a low-latency, sparsity-aware stream compression approach for first-level trigger systems in collider experiments. Our method is implemented as a fully configurable, open-source hardware generator in Chisel, optimized for FPGAs. Evaluation across multiple configurations demonstrates robust performance, and integration into the GNN-ETM within the Belle~II first-level trigger confirms feasibility. In this use case, we achieve a reduction of the computational load for the subsequent GNN inference by a factor of 324 in comparison to a naive approach at a latency overhead of below $60\,\text{ns}$.
To conclude, our results demonstrate that sparsity-aware compression enables efficient machine learning on FPGAs. Our source code is available on GitHub~\cite{zenodo:18644386}.

\begin{figure}[ht]
    \centering
    \small
     \begin{tikzpicture}
        \pgfplotsset{
        every axis legend/.append style={
            at={(0.55,1.25)},
            anchor=north west,
            legend columns = -1}}
    
        \begin{groupplot}[
            group style={
                group size=3 by 1,
                x descriptions at=edge bottom,
                vertical sep=40pt,
                horizontal sep=20pt,
                ylabels at=edge left,
                yticklabels at=edge left,
                xlabels at=edge bottom,
                xticklabels at=edge bottom
                },
            ymin=0,
            ymax=15,
            ytick={0,5,10,15},
            yticklabels={0~\%,5~\%,10~\%,15~\%},
            minor y tick num=5,
            symbolic x coords={LUT,FF,CLB}
        ]
    
        \nextgroupplot[
            title={$N_I = 64$},
            title style={at={(0.5,1.00)},anchor=south,yshift=-0.1},
            width= .4\textwidth,
            height= .35\textwidth,
            major x tick style = transparent,
            ybar=1pt,
            bar width=8pt,
            ymajorgrids = true,
            xtick = data,
            scaled y ticks = false,
            enlarge x limits=0.2,
            enlarge y limits=0.1,
            line width=1pt,
            legend columns=1,
            legend style={
                at={(0.05,0.95)},
                anchor=north west},
            legend image code/.code={\draw[#1] (0cm,-0.1cm) rectangle (0.2cm,0.1cm);}
    ]
            
            \addplot[style={white,fill=KITblue,mark=none}]
                coordinates {(FF, 0.58) (LUT,0.91) (CLB,1.46)};
    
            \addplot[style={white,fill=KITpurple,mark=none}]
                coordinates {(FF, 0.54) (LUT,1.34) (CLB,1.84)};
    
            \addplot[style={white,fill=KITforestgreen,mark=none}]
                coordinates {(FF, 0.48) (LUT,2.75) (CLB,3.22)};
    
            \addlegendentry{$N_O = 2$}
            \addlegendentry{$N_O = 4$}
            \addlegendentry{$N_O = 8$}

        \nextgroupplot[
            title={$N_I = 128$},
            title style={at={(0.5,1.00)},anchor=south,yshift=-0.1},
            width= .4\textwidth,
            height= .35\textwidth,
            major x tick style = transparent,
            ybar=2pt,
            bar width=8pt,
            ymajorgrids = true,
            xtick = data,
            line width=1pt,
            scaled y ticks = false,
            enlarge x limits=0.2,
            enlarge y limits=0.1]
            
            \addplot[style={white,fill=KITblue,mark=none}]
                coordinates {(FF, 1.18) (LUT,1.87) (CLB,3.03)};
    
            \addplot[style={white,fill=KITpurple,mark=none}]
                coordinates {(FF, 1.12) (LUT,2.82) (CLB,3.58)};
    
            \addplot[style={white,fill=KITforestgreen,mark=none}]
                coordinates {(FF, 1.06) (LUT,5.88) (CLB,5.97)};
        
        \nextgroupplot[
            title={$ N_I = 256$},
            title style={at={(0.5,1.00)},anchor=south,yshift=-0.1},
            width= .4\textwidth,
            height= .35\textwidth,
            major x tick style = transparent,
            ybar=1pt,
            bar width=8pt,
            ymajorgrids = true,
            xtick = data,
            line width=1pt,
            scaled y ticks = false,
            enlarge x limits=0.2,
            enlarge y limits=0.1]
            
            \addplot[style={white,fill=KITblue,mark=none}]
                coordinates {(FF, 2.39) (LUT,3.86) (CLB,4.67)};
    
            \addplot[style={white,fill=KITpurple,mark=none}]
                coordinates {(FF, 2.29) (LUT,6.38) (CLB,6.87)};
    
            \addplot[style={white,fill=KITforestgreen,mark=none}]
                coordinates {(FF, 2.22) (LUT, 12.28) (CLB, 15.05)};
                
        \end{groupplot}
        
        \end{tikzpicture}
    \caption{Relative system resource utilization for various module configurations after out-of-context synthesis, place and route with AMD Vivado 2024.2 for the AMD Ultrascale XCVU190.}
    \label{fig:utilization}
\end{figure}
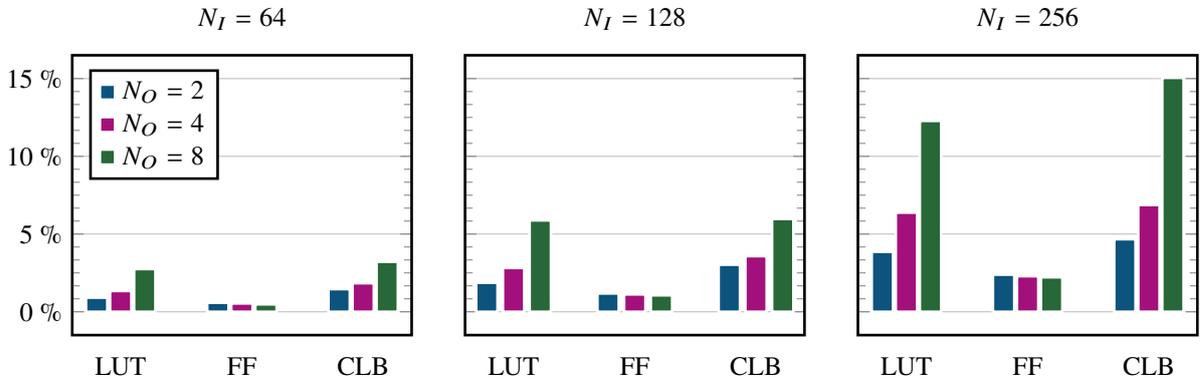

\appendix
\acknowledgments
We kindly thank Thomas Lobmaier for his support in generating a visual representation of the Belle II Event Display.


\bibliographystyle{JHEP}
\bibliography{biblio.bib}



\end{document}